\documentclass{aa}  
\usepackage{natbib}
\bibpunct{(}{)}{;}{a}{}{,} 
\usepackage{graphicx}
\usepackage{txfonts}
\begin{document}

   \title{Signatures of warm dark matter in the cosmological density fields extracted using Machine Learning}

   \subtitle{}

   \author{Ander Artola
          \inst{1}
          \and
          Sarah E.~I.~Bosman\inst{1}\fnmsep\inst{2}\thanks{\email{bosman@thphys.uni-heidelberg.de}}
          \and
          Prakash Gaikwad\inst{2}
          \and
          Frederick B.~Davies\inst{2}
          \and
          Fahad Nasir\inst{2}
          \and
          Emanuele P.~Farina\inst{3}
		\and
          Klaudia Protu\v{s}ová\inst{1}
			\and
          Ewald Puchwein\inst{4}
		\and
			Benedetta Spina\inst{1}
          }

   \institute{Institute for Theoretical Physics, Heidelberg University, Philosophenweg 12, D-69120, Heidelberg, Germany
         \and
             Max-Planck-Institut f\"ur Astronomie, K\"onigstuhl 17, 69117 Heidelberg, Germany
        \and 
            Gemini Observatory, NSF's NOIRLab, 670 North A'ohoku Place, Hilo, HI 96720, USA
        \and
            Leibniz-Institut f\"ur Astrophysik Potsdam, An der Sternwarte 16, D14482 Potsdam, Germany
             }

   \date{}

  \abstract
   {}
   {We aim to construct a machine-learning approach that allows for a pixel-by-pixel reconstruction of the intergalactic medium (IGM) density field for various warm dark matter (WDM) models using the Lyman-$\alpha$ forest. With this regression machinery, we constrain the mass of a potential warm dark matter particle from observed Lyman-$\alpha$ sightlines directly from the density field.}
   {We design and train a Bayesian neural network on the supervised regression task of recovering the optical depth-weighted density field $\Delta_\tau$ as well as its reconstruction uncertainty from the Lyman-$\alpha$ forest flux field. We utilise the \texttt{Sherwood-Relics} simulation suite at $4.1\leq z\leq 5.0$ as the main training and validation dataset. Leveraging the density field recovered by our neural network, we construct an inference pipeline to constrain the warm dark matter particle masses based on the probability distribution function of the density fields. We implement a rudimentary approach to marginalise over the IGM temperature by selecting the \texttt{Sherwood-Relics} run which best fits the target data, which in our analysis is always the coldest available model.}
   {We find that our trained Bayesian neural network can accurately recover within a $1\sigma$ error $85\%$ of the density fields from a validation simulated dataset that encompasses multiple warm dark matter and thermal models of the IGM. When predicting on simulated Lyman-$\alpha$ skewers generated using the alternative hydrodynamical code \texttt{Nyx}, not included in the training data, we find a $1\sigma$ accuracy rate $\geq 75\%$. We consider 2 samples of observed Lyman-$\alpha$ spectra from the UVES and GHOST instruments, at $z=4.4$ and $z=4.9$ respectively and fit the density fields recovered by our Bayesian neural network to constrain warm dark matter masses. We find lower bounds on the warm dark matter particle mass of $m_{\mathrm{WDM}} \gtrsim 3.8$ KeV and $m_{\mathrm{WDM}} \gtrsim 2.2$ KeV at $2\sigma$ confidence, respectively. We note that these constraints may be weakened by a more thorough treatment of the IGM thermal state.}
   {Using machine learning, we are able to match current state-of-the-art WDM particle mass constraints using up to $\sim 40$ times less observational data than Markov Chain Monte Carlo techniques based on the Lyman-$\alpha$ forest power spectrum.}


   \maketitle
%

\section{Introduction}
Many aspects of the large-scale structure of the Universe can be understood by including a \emph{cold dark matter} (CDM) component that represents $\sim 26 \%$ of the critical density \citep{2014A&A...571A..16P}. In the standard cosmological model, $\Lambda$CDM, dark matter is an important ingredient for structure formation in the Early Universe. However, its precise nature remains an outstanding problem both in cosmology and particle physics. With more exotic dark matter (DM) candidates, such as primordial black holes, heavily constrained, it is likely that DM consists of some undiscovered elementary particle(s) produced early in the history of the universe \citep{Villanueva_Domingo_2021}. 

On large scales, the predictions of $\Lambda$CDM have been amply tested and are in good agreement with observations \citep{Dalal2002, VanWaerbeke2004, Eisenstein_2005}. In contrast, on scales smaller than $\sim 10$ kpc, potential tensions between CDM predictions and observations might exist, including the ``core-cusp'' problem related to the DM density profile in halos, or the ``too big to fail'' problem linked to the number density of high-luminosity satellites in sub-halos \citep{moore_evidence_1994, 10.1111/j.1745-3933.2011.01074.x, Weinberg_2015}. Even if the inclusion of complex baryonic feedback processes can alleviate the aforementioned potential discrepancies, alternative models to CDM are worth exploring \citep{vogelsberger_properties_2014}.  

From a cosmological standpoint, many DM models are distinguished based on their velocity dispersion, which defines the average distance travelled by DM particles. This distance is referred to as the comoving free-streaming length $\lambda_{\text{FS}}$. On scales smaller than $\lambda_{\text{FS}}$, the velocity dispersion of dark matter particles suppresses the gravitational clustering of matter, and consequently, structure formation. By definition, the free-streaming length for CDM particles is negligible at the scales of cosmological structure formation. Conversely, hot dark matter models, such as light neutrinos, smoothe out gravitational clustering even at galaxy cluster scales, leading to tight constraints on such models \citep{Hannestad_2004}. In this work, we focus on a set of intermediate models known as warm dark matter (WDM), with typical masses in the range scale $\sim 1$ KeV \citep{Viel_2005}. For reference, a thermal relic\footnote{Thermal relics are models where the DM particles were coupled to the primordial plasma in the Early Universe. In contrast, non-thermal relics include models without such thermal equilibrium.} WDM model with a particle mass of 1 KeV has an associated free streaming length of $\lambda_{\text{FS}} \sim 0.3$ Mpc. For these WDM models, the scale at which clustering suppression occurs is probed by the Lyman-$\alpha$ forest, which consists of absorption features observed in the spectra of quasars. These features are produced by the absorption of redshifted Lyman-$\alpha$ photons as they traverse the low-density IGM
\citep{1965ApJ...142.1633G, Fan2002}.

The Lyman-$\alpha$ forest is a rich source of information. Previous works have successfully used the Lyman-$\alpha$ forest to constrain the thermal parameters of the IGM \citep{Boera_2019, Gaikwad_2021, wolfson2023forecastingconstraintshighzigm}. The common theme of these previous approaches has been to use summary statistics of the Lyman-$\alpha$ flux (such as the power spectrum, probability distribution function, etc) to directly constrain thermal or reionisation parameters. Numerous efforts have also been made to constrain WDM models directly at the Lyman-$\alpha$ flux level \citep{Ir_i__2017, Villasenor_2023, Ir_i__2024}.

With the recent disruption of Machine Learning (ML), deep learning techniques have become increasingly popular in analysing simulations and real data. Recently, ML approaches have shown great success in extracting information from the Lyman-$\alpha$ forest \citep{Nayak_2024,maitra2024parameterestimationlyalphaforest}. In this work, we build on previous efforts to recover the IGM gas conditions from Lyman-$\alpha$ sightlines \citep{nasir2024}.

In this paper, we present a new machine learning-based approach to constrain WDM candidates based on inference directly at the density field level. In Section \ref{sec: nn} we introduce a Bayesian neural network architecture that can do regression and predict an optical depth-weighted density field $\Delta_\tau$, as well as the uncertainity in the reconstruction, from a given Lyman-$\alpha$ skewer. In Section \ref{sec: sherwood} we present the training data over which we optimise the neural network parameters. We use the \texttt{Sherwood-Relics} simulation suite as the main dataset, both for training and validation and include independent data generated by the \texttt{Nyx} code as a last validation step. Section \ref{sec: inference method} presents the statistical inference method that we use to obtain WDM constraints from a sample of Lyman-$\alpha$ skewers by leveraging the trained neural network predictions. Section \ref{sec: data} is devoted to presenting the observational data that will be later used,  in Section \ref{sec:results}, to obtain our final results. Throughout this paper, comoving distances are denoted with cMpc, and the following best-fit $\Lambda$CDM cosmological parameters from $Planck+WP+highL+BAO$ are assumed: $\Omega_m=0.308$, $\Omega_\Lambda=0.692$, $h=0.678$, $\Omega_b=0.0482$, $\sigma_8=0.829$, $n=0.961$ and a primordial helium fraction of $Y=0.24$ \citep{2014A&A...571A..16P}. These ``intermediate'' \textit{Planck} cosmological parameters, which are consistent with the final ones \citep{Planck_final}, are used for consistency with the fiducial runs of the \texttt{Sherwood-Relics} simulation.

\section{Methods}\label{sec: methods}

\subsection{Bayesian neural network architecture}\label{sec: nn}
We leverage the implementation of Bayesian neural networks in \cite{nasir2024} to train a machine learning model in a supervised regression task to predict the $\Delta_\tau$ optical depth-weighted density field from a Lyman-$\alpha$ flux skewer. To quantify the model's confidence in a prediction, we utilise Bayesian networks, which predict a full posterior probability distribution at each pixel for the target density field \citep{Jospin_2022}. The uncertainties associated with these machine learning predictions can then be quantified and propagated through the rest of the statistical analysis, as described in Section \ref{sec: inference method}.
The code implementing this architecture is publicly available\footnote{\url{https://github.com/nicenustian/bh2igm}} and written using the \texttt{TensorFlow} and \texttt{TensorFlow-Probability} frameworks \citep{tensorflow2015-whitepaper}.

\begin{figure*}[t]
    \centering
    \includegraphics[width=0.9\textwidth]{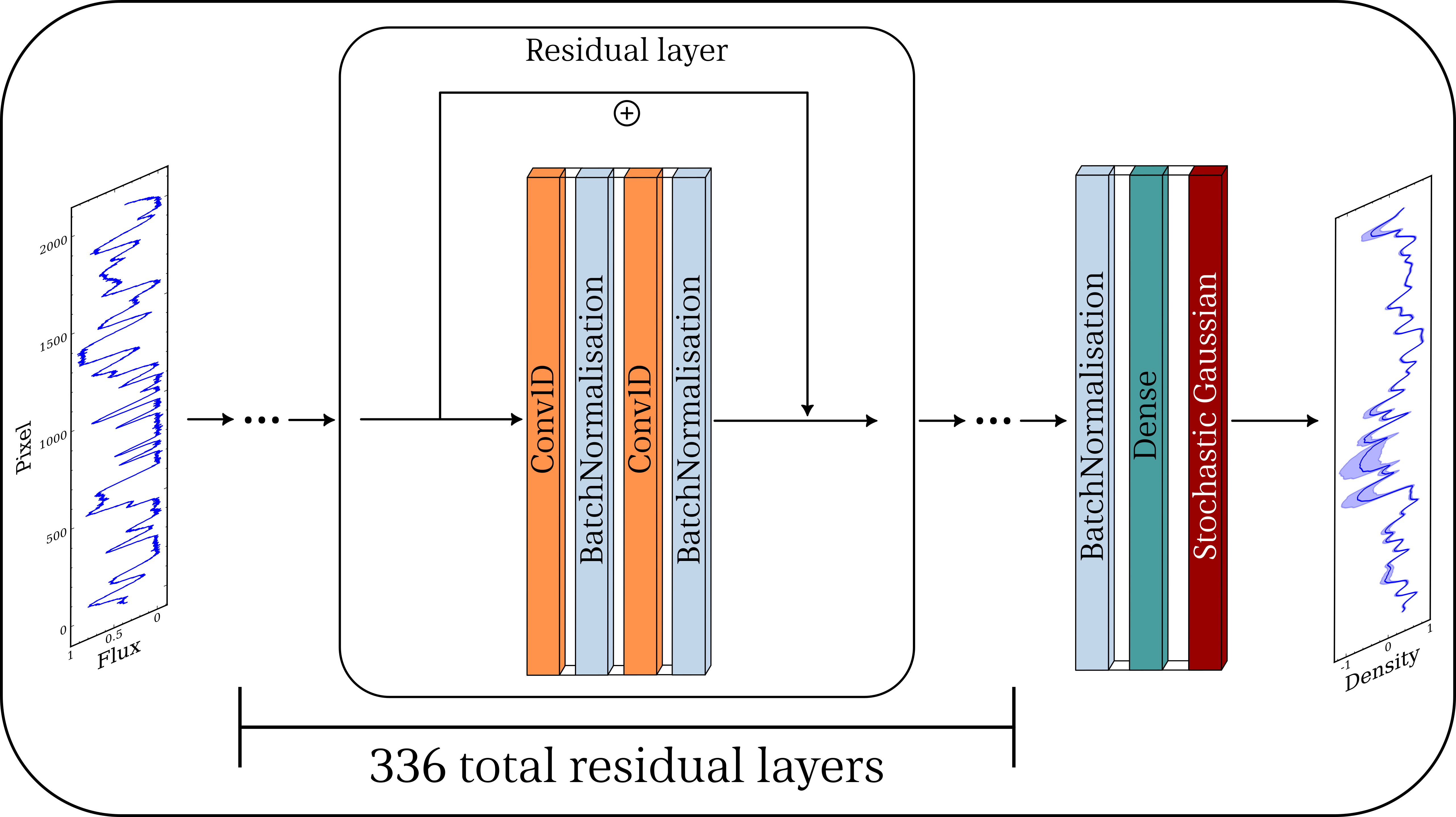}
    \caption{Network topology for the fiducial hyper-parameter search at $z=4.4$ using \texttt{OPTUNA} and trained on the \texttt{Sherwood-Relics} simulation suite with varied thermal history and WDM masses, as shown in Table \ref{tab: Sherwood}. The neural network takes as input a one-dimensional flux sightline and produces a normal distribution at each $\Delta_\tau$ pixel as output. The basic building blocks are 336 residual layers, which include convolutional and batch normalisation layers.}
              \label{fig: NN_archi}%
\end{figure*}

Figure \ref{fig: NN_archi} shows a schematic representation of the fiducial neural network architecture trained at $z=4.4$ on the \texttt{Sherwood-Relics} suite (see Section \ref{sec: sherwood} for the details on the training data) with different thermal histories and WDM masses according to Table \ref{tab: Sherwood}.

The neural network takes as input a one-dimensional Lyman-$\alpha$ sightline and produces as output a one-dimensional $\Delta_\tau$ field, defined in velocity space as:
\begin{equation}
\Delta_{\tau,i} = \frac{1}{\tau_i} \displaystyle\sum_j \Delta_j N_{\text{HI}, j} \phi_\alpha(\Delta v_{ij}),
\end{equation}
where $\tau_i = \sum_j N_{\text{HI}, j} \phi_\alpha(\Delta v_{ij})$, $N_{\text{HI}, j}$ is the column density of neutral hydrogen at pixel $j$, $\phi_\alpha(v)$ is the cross-section of Lyman-$\alpha$ absorption, $\Delta v_{ij}$ is the velocity separation between pixels $i$ and $j$, and $\Delta_j$ is the density (e.~g.~\citealt{Schaye99}). 
The neural network generates a mean and standard deviation for each density pixel, modelled as a Gaussian. The basic building blocks for the fiducial architecture are residual blocks. They include two one-dimensional convolutional layers with a fixed filter size of 3 pixels and two batch normalisation layers with a skipping connection. The Bayesian nature of the network is implemented as a stochastic activation layer as the final component: a dense layer produces the mean and standard deviation that are used as the parameters for each Gaussian density pixel. We train the network to maximise the data likelihood with respect to the output distribution. Hence, we use as loss function the negative log-likelihood, defined as
\begin{equation}
    -\log\mathcal{L}=\frac1N\sum_\mathrm{i}\left((Y_\mathrm{i}-\mu_i)^2/\sigma_\mathrm{i}^2+\log(\frac1{\sigma_\mathrm{i}^2})\right),
\end{equation}
where $N$ is the number of pixels in a sightline, $\mu_i$ is the mean predicted $\Delta_\tau$ value at pixel $i$, $\sigma_i$ the standard deviation, and $Y_i$ the true $\Delta_\tau$ value.
We fine-tune the network hyper-parameters using \texttt{OPTUNA}, a Python API for automatic hyper-parameter optimisation based on the Tree Parzen Estimation algorithm \citep{optuna, TPE}. The parameters optimised are the learning rate, the architecture type, the batch size, the number of layers and the number of convolutional filters per layer.
We summarise their optimal values at $z=4.4$ in Table \ref{tab: nn_archi}. Note that the hyper-parameters should be fine-tuned even if the training data is slightly changed. For instance, training with data at a different redshift can require a different number of convolutional filters to reach optimal performance. Additionally, note that this optimisation is done purely on a performance basis. As a consequence, it is well-known that such networks tend to be over-parameterised \citep{Fang_2021}.

\begin{table}
      \caption[]{Fiducial hyper-parameter values for the neural network architect obtained using \texttt{OPTUNA} at $z=4.4$ when training on the \texttt{Sherwood-Relics} suite presented in Section \ref{sec: sherwood}.}
         \label{tab: nn_archi}
     $$ 
         \begin{array}{p{0.3\linewidth}c}
            \hline
            \noalign{\smallskip}
            Parameter      & \mathrm{Optimal\ value\ at }\ z=4.4 \\ 
            \noalign{\smallskip}
            \hline
            \noalign{\smallskip}
            Type & \mathrm{Residual}  \\
            Learning rate & 0.00494  \\
            Batch size & 32  \\
            Layer layout     & [1, 2, 4, 4]         \\
             Filter per layer & [16, 32, 32, 32]  \\
            \noalign{\smallskip}
            \hline
         \end{array}
     $$ 
   \end{table}

\subsection{Simulated training data}\label{sec: sherwood}
The vast majority of supervised regression tasks require a large and varied quantity of labelled training data to ensure model convergence and appropriate generalisation capabilities. In this work, we leverage the \texttt{Sherwood-Relics} simulation suite, a set of high-resolution cosmological hydrodynamical simulations specially designed to resolve the Lyman-$\alpha$ forest \citep{Bolton_2016, Puchwein23, Ir_i__2024}. The \texttt{Sherwood-Relics} simulations are based on a modified version of the parallel Tree-PM Smoothed Particle Hydrodynamics code \texttt{P-GADGET-3}, which is in turn based on the publicly-released \texttt{GADGET-2} code \citep{Springel_2005}. 
We consider the set of runs as described in Table \ref{tab: Sherwood}. We use runs with a 20h$^{-1}$cMpc box length, $1024^3$ dark matter and gas particles and a uniform UVB background as described in \cite{Puchwein2019, Puchwein23}. The runs have the same random initial seed but vary the WDM mass parameter. The fiducial run includes CDM, and modified runs include WDM masses of \hbox{\{2, 3, 4, 8, 12 \}} KeV. 
Since a hotter IGM temperature can also suppress the power spectrum of the Lyman-$\alpha$ forest on small scales, we wish to break its degeneracy with the WDM particle mass. 
We therefore also consider runs with varied thermal histories, labelled \texttt{ref}, \texttt{hot} and \texttt{cold} which results in a variation of the temperature at mean density in the IGM, $T_0$. At $z=4.6$, the \texttt{cold} run has $T_0=6598$ K, the \texttt{hot} run has $T_0=13957$ K and the \texttt{ref} run has $T_0=10066$ K. For more details, see \cite{Puchwein23, Ir_i__2024}. With such varied training data comprising a total of 18 simulation boxes, we aim to build a neural network agnostic to the exact thermal and WDM parameters. Note, however, that we do not include runs that vary the slope $\gamma$ of the temperature-density relationship $T=T_0\Delta^{\gamma-1}$; see \cite{nasir2024} for a discussion of the inclusion of such runs in the performance of the neural network.

\begin{table}
      \caption[]{List of the \texttt{Sherwood-Relics} runs used in the work. For more details, see \cite{Ir_i__2024}.
      All box sizes are 20h$^{-1}$. The table shows the temperature at mean density in the IGM $T_0$ at redshift $z=4.6$ and the set of WDM masses included. We work with the inverse WDM mass and consider 0 to correspond to the CDM reference run.}
         \label{tab: Sherwood}
     $$ 
         \begin{array}{p{0.3\linewidth}cc}
            \hline
            \noalign{\smallskip}
            Run      &  T_0 {[\mathrm{K}]}\ (z=4.6) & \mathrm{WDM} [\mathrm{KeV}^{-1}] \\ 
            \noalign{\smallskip}
            \hline
            \noalign{\smallskip}
            L20-ref & 10066 &\{0,\frac{1}{2}, \frac{1}{3}, \frac{1}{4}, \frac{1}{8}, \frac{1}{12} \}     \\
            L20-ref-hot           & 13957  &\texttt{"}\\
            L20-ref-cold     & 6598  &       \texttt{"}     \\
            \noalign{\smallskip}
            \hline
         \end{array}
     $$ 
\end{table}

\begin{figure*}[t]
   \centering
   \includegraphics[width=0.99\textwidth]{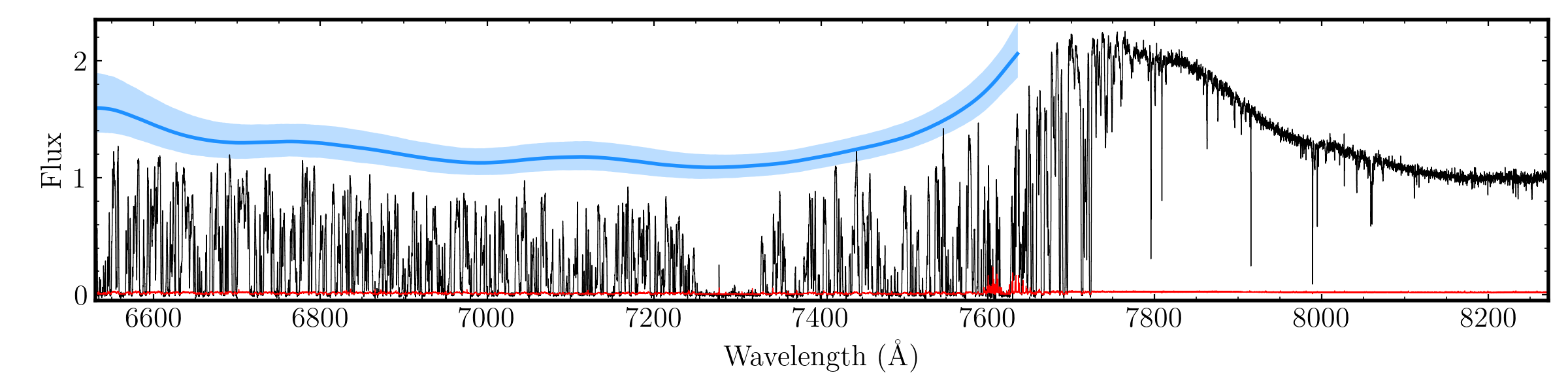}
      \caption{Spectrum of J0306+1853 observed with the GHOST instrument. We show in blue the PCA reconstruction of the emission continuum based on the spectrum redward of the Lyman-$\alpha$ line and its $1\sigma$ uncertainty.}
         \label{fig: ghost spectrum}
\end{figure*}

We obtain 5000 sightlines with 2048 pixels from each run at redshift bins of $\Delta z=0.1$ from $z=4.1$ to $z=5.0$, including the Lyman-$\alpha$ optical depth and the neutral hydrogen overdensity $\Delta$. As described in \cite{nasir2024}, we form pairs of Lyman-$\alpha$ flux skewers and optical depth-weighted densities $\Delta_\tau$, which will be the central data in the rest of this work. We normalise the Lyman-$\alpha$ flux by re-scaling the Lyman-$\alpha$ optical depth to match the mean observed flux \citep{Becker2013, Bosman2018, Bosman2022}.
During the training of the model, the flux skewers are processed on the fly with a user-specified noise level, instrumental resolution (implemented as a convolution with a Gaussian kernel), binning to a desired number of pixels, and by applying random translations. This gives enough flexibility to the exact characteristic of the data,  allowing us to use our machinery on diverse real observations of the Lyman-$\alpha$ forest and preventing over-fitting.

During training, $80\%$ of the \texttt{Sherwood-Relics} skewers are used for updating the model's weights, and the rest $20\%$ are used as validation. The model's weights are only updated if the performance on the validation split increases.

Additionally, we use an independently simulated set of Lyman-$\alpha$ skewers and $\Delta_\tau$ fields generated using the \texttt{Nyx} simulation code as testing data \citep{Almgren_2013, Lukic15}. With this, we aim to validate the model robustness on data generated using different hydrodynamical solvers, physical prescriptions and thermal histories. Successful prediction on this data set will be a strong sign of adequate generalisation capabilities by the trained neural network. We use 3 CDM Nyx runs in this testing step, with the same box length of 20h$^{-1}$cMpc as the \texttt{Sherwood-Relics} runs, and different thermal histories that vary the reionisation redshift, defined as the redshift when the neutral hydrogen fraction is lower than $10^{-3}$. We label the 3 runs by their reionisation redshift as $\{\mathrm{zre6}, \mathrm{zre7}, \mathrm{zre8}\}$. From each run, we extracted 5000 validation skewers.

\begin{table}
      \caption[]{The six sightlines from the SQUAD sample used in this work.}
         \label{tab: squad dr1}
     $$ 
         \begin{array}{p{0.4\linewidth}cc}
            \hline
            \noalign{\smallskip}
            \text{SQUAD DR1 name} & z_\text{em} & \text{SNR} \\ 
            \noalign{\smallskip}
            \hline
            \noalign{\smallskip}
            J004054 & 4.976 & 33      \\
			J021043 & 4.65 & 25 \\
			J025019 & 4.77 & 12 \\
			J030722 & 4.728 & 50 \\
			J033829 & 5.032 & 14 \\
			J145148 & 4.763 & 100 \\
            \noalign{\smallskip}
            \hline
         \end{array}
     $$ 
   \end{table}
   
\subsection{Statistical inference on the warm dark matter}\label{sec: inference method}
We implement a statistical inference pipeline to recover WDM masses from the reconstructed $\Delta_\tau$ skewers. We consider a target set of $N$ observed Lyman-$\alpha$ forest sightlines centred at $z=z_\mathrm{obs.}$, each with potentially different instrumental resolution, binning, and noise. We train our neural network on the \texttt{Sherwood-Relics} suite at $z=z_\mathrm{obs.}$ with the same instrumental resolution, binning, and noise for each independent sightline. We then consider the predicted mean values for the $\Delta_\tau$ field by the neural network, $\{ \mu(i,j) \}_{i,j}$, where index $i=0,..., N$ labels the sightline and $j$ labels the pixel. We take the $\Delta_\tau$ Probability Distribution Function (PDF) as the target summary statistic. We compute the observed PDF by computing the histogram of the aggregated predicted fields $\{ \Delta_\tau(i,j) \}_{i,j}$, which we label as $\mathrm{PDF}_\mathrm{obs.}$, where now $j$ labels the bin index. As we will describe in Section \ref{sec:results}, the pixels of a sightline where the absorption is saturated are dominated by noise, and the neural network fails to give precise predictions for the density field. When computing the PDFs, we neglect the pixels that are saturated by applying a mask whenever the flux falls below the quantity $3/\mathrm{SNR}$, where $\mathrm{SNR}$ is the median signal-to-noise ratio of the sightline. 

We estimate two different independent uncertainties on the $\mathrm{PDF}$. Firstly, we estimate the uncertainties associated with the finite sample variance by bootstrapping the $N$ observations 1000 times and calculating the scatter in $\mathrm{PDF}(j)$. Secondly, we follow \cite{nasir2024} and calculate the uncertainty associated with the Bayesian neural network predictions. For that purpose, we estimate a covariance matrix $\Sigma$ of the predictions residuals $r_i$ for the reference \texttt{Sherwood-Relics} run,
\begin{equation}\label{eq:residuals}
    r_i=\frac{\Delta_{\tau, i}-\mu_i}{\sigma_i},
\end{equation}
where $\sigma_i$ are the machine-learning-predicted uncertainties at pixel $i$.
We sample 1000 times each recovered $\Delta_\tau$ skewer using a multivariate normal with covariance $\Sigma$. We then include the specific pixel uncertainties $\sigma_i$ for each resampled prediction. Finally, we take the scatter in the $\mathrm{PDF}$ from the samples as the second independent uncertainty on the $\mathrm{PDF}_\mathrm{obs.}$. We denote the total error on $\mathrm{PDF}_\mathrm{obs.}(j)$ as $\varepsilon(j)$.

We now fit the observed statistic, $\mathrm{PDF}_\mathrm{obs.}$, to each of the model statistic obtained from the \texttt{Sherwood-Relics} suite with matching noise and resolution. For each \texttt{Sherwood-Relics} model with thermal history $t\in\{\texttt{ref}, \texttt{hot}, \texttt{cold} \}$ and WDM mass $m$, we compute the forward-modelled $\mathrm{PDF}$ across all skewers and denote it by $\mathrm{PDF}(j,t,m)$, where $j$ labels again the bin. To fit the model statistics to the observed statistic, we use a simple $\chi^2$ metric defined as
\begin{equation}\label{eq:chi}
    \chi^2(t,m)=\sum_j \frac{\left(  \mathrm{PDF}(j,t,m)-\mathrm{PDF}_\mathrm{obs.}(j)  \right)^2}{\left(\varepsilon(j)\right)^2}.
\end{equation}
Since in this work, we are only interested in one-parameter constraints on the WDM mass and not on the thermal history or parameters, we define the best-fit model $(t,m)$ as
\begin{equation}\label{eq:best_fit_def}
    \mathrm{min}_{t,m} \, \, \chi^2(t,m).
\end{equation}
As we will discuss later in section \ref{sec:results}, all the best fit WDM mass models are typically $0$ KeV$^{-1}$, which only leads to lower limits on the WDM mass. Hence, equation \ref{eq:best_fit_def} is equivalent to only considering the thermal model that minimises the $\chi^2$ value on the CDM model. In this manner, we reduce the problem to a one-dimensional fit. With the best-fit values for $(t,m)$, we can obtain constraints on the WDM mass $m$ by calculating the confidence regions on the fit, which are given by
\begin{equation}\label{eq: confidence_region}
    \chi^2-\chi^2_\mathrm{min.}=\{1,4\},
\end{equation}
for $1\sigma$ and $2\sigma$, respectively (e.g.~\citealt{Avni76}).

\begin{figure*}[t]
    \centering
    \includegraphics[width=0.99\textwidth]{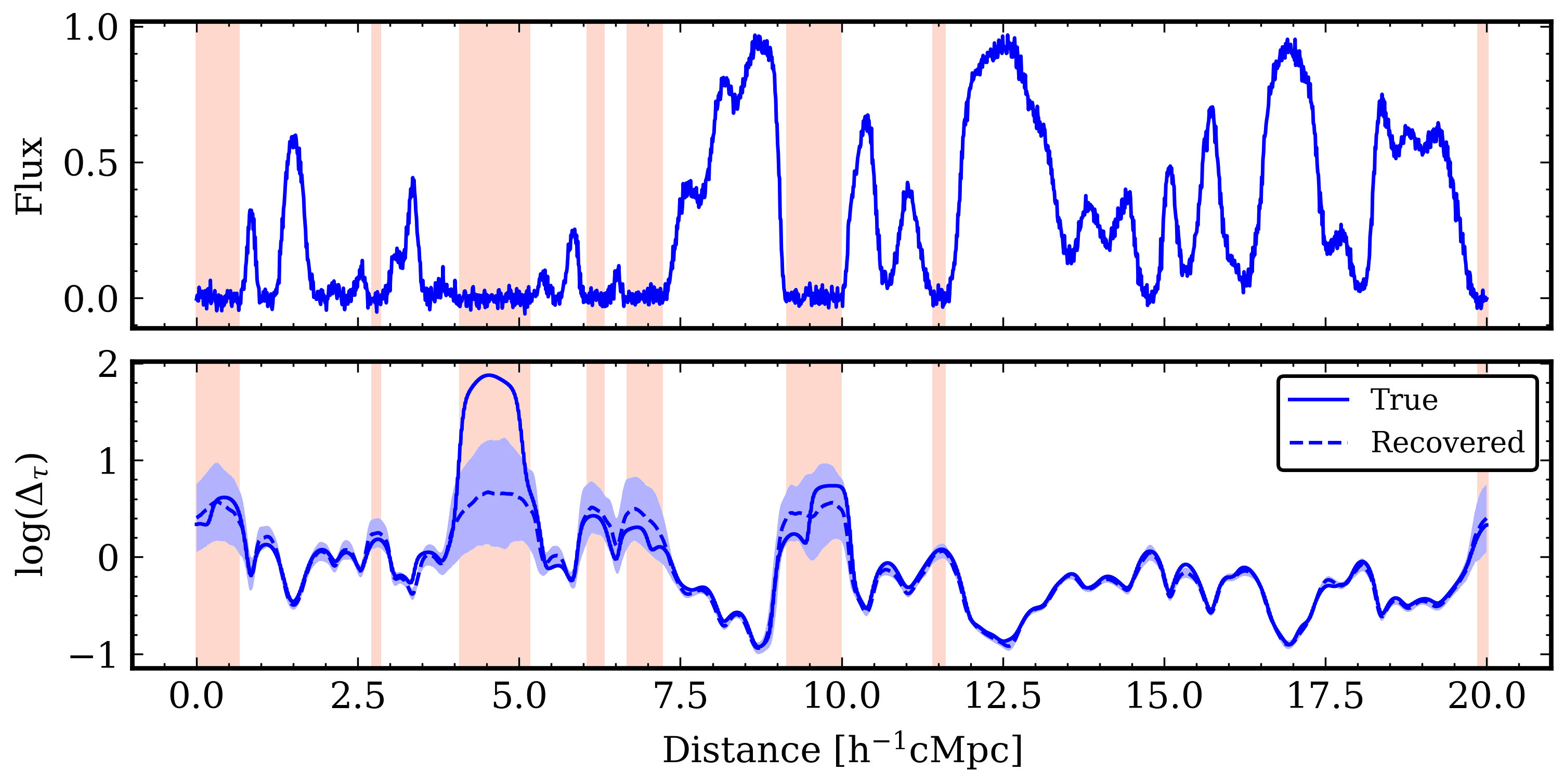}
    \caption{Our trained neural network on the \texttt{Sherwood-Relics} data predicting the $\Delta_\tau$ field on a 20h$^{-1}$cMpc CDM skewer from the \texttt{Nyx} zre6 run. The top panel shows the input Lyman-$\alpha$ flux with added Gaussian noise (SNR $=50$) and an instrumental resolution of $6$ km\ s$^{-1}$. The bottom panel shows the machine learning prediction, with the true $\Delta_\tau$ field as a solid curve, and the mean prediction as a dashed curve. The blue region shows the $1\sigma$ uncertainty predicted by the Bayesian neural network. Note that the neural network has not been trained on any data generated using the \texttt{Nyx} code. The network's predictions only differ significantly from the truth over regions with no flux and therefore no information; such regions are masked from our analysis (orange shading).}
              \label{fig: cdm_recovered}%
\end{figure*}

   \begin{figure}[h]
   \centering
   \includegraphics[width=\columnwidth]{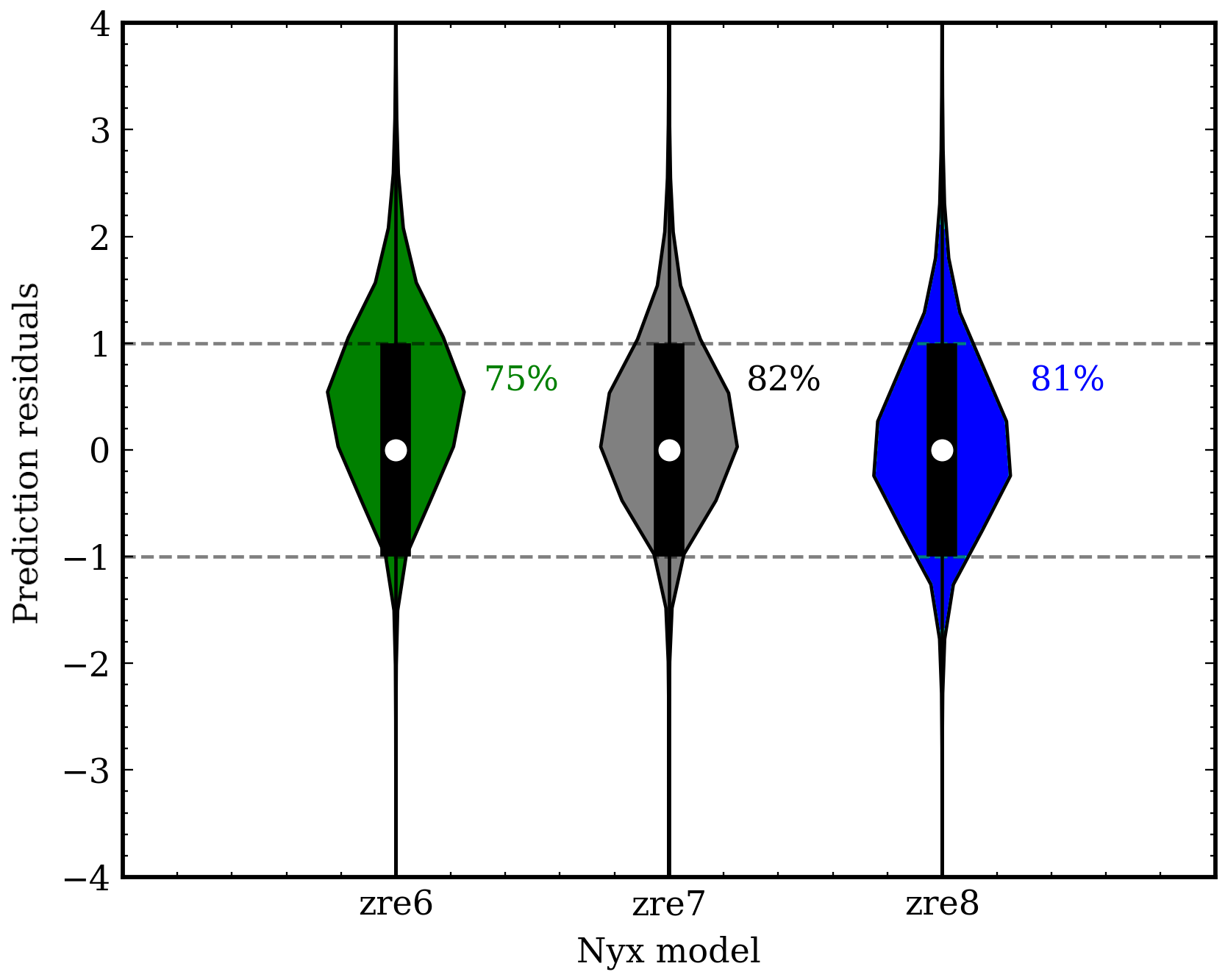}
      \caption{Prediction residuals, as defined in equation \ref{eq:residuals}, for the 3 different validation \texttt{Nyx} runs. The corresponding percentages indicate the pixel fraction that is correctly recovered within $1\sigma$.}
         \label{fig: violin Nyx}
   \end{figure}

\section{Data}\label{sec: data}
We apply our density recovery and WDM mass inference pipeline to two observed samples. Firstly, a set of 6 observed quasar sightlines from the SQUAD DR1 survey \citep{Murphy_2018}. Secondly, to a single spectrum from the newly-commissioned Gemini High-resolution Optical SpecTrograph (GHOST, \citealt{GHOST1,GHOST2}). With these two samples, we wish to test the real performance and stability of our method on observational data with very different properties, but which still resolve the Lyman-$\alpha$ forest.

The SQUAD DR1  data consists of 6 Lyman-$\alpha$ sightlines of size 20h$^{-1}$cMpc with varied SNR (see Table \ref{tab: squad dr1}), observed with the Ultraviolet and Visual Echelle Spectrograph (UVES, \citealt{UVES}) on the European Southern Observatory's Very Large Telescope, which has an average resolution of $\mathrm{FWHM}\approx 6$ km s$^{-1}$. We consider sightlines centred at $z=4.4$ for this specific application and 1039 pixels long.

We also consider a Lyman-$\alpha$ skewer obtained from the GHOST instrument, which corresponds to the ultra-luminous quasar J0306+1853 with emission redshift $z=5.363$ \citep{Wang_2015}. The data were obtained during a commissioning run of the instrument during which the new spectrograph was slightly out of focus, leading to a resolution of $R \sim 30000$, less than the instrument's nominal performance of $R\sim55000$ (2 hours of integration; PID GS-2023B-FT-204; P.~I.~Bosman). 
For GHOST, we focus our attention on a slightly higher redshift, $z=4.9$, above which  a Damped Lyman-$\alpha$ system (DLA) at $\sim 1150$ \textup{~\AA} occurs. We extract 2 skewers of length 20h$^{-1}$cMpc, one terminating at $z=4.9$ and the other beginning at the same redshift, and consider them to be independent in order to run the neural network predictions. We compare to the \texttt{Sherwood-Relics} snapshot at $z=4.9$. 
We performed a continuum reconstruction over the Lyman-$\alpha$ forest with a PCA technique based on the spectrum to the right side of the Lyman-$\alpha$ line, following the method of \citet{Bosman21}. The spectrum and the reconstruction are shown in Figure \ref{fig: ghost spectrum}. 

Due to limitations in the fluxing calibration of the instrument, the continuum normalisation in the SQUAD sample is defined in such a manner as the peaks of the Lyman-$\alpha$ forest transmission spikes reach to the $100\%$ of the continuum. In reality the peaks should only reach to $C_\mathrm{corr}\approx 80\%$ of the unabsorbed continuum \cite{Bolton_2016}, as is the case in both the GHOST data as well as the \texttt{Sherwood-Relics} and \texttt{Nyx} skewers. We post-process the skewers to match the normalisation as defined in the SQUAD sample whenever the data is utilised.

\section{Results}\label{sec:results}
In this section, we discuss the performance of the trained neural network on the validation split of the \texttt{Sherwood-Relics} data, as well as on the \texttt{Nyx} runs used as validation. We also test the constraining potential of the inference pipeline discussed in Section \ref{sec: inference method} by first using simulated skewers, and then using the real observed sightlines presented in Section \ref{sec: data}.

   \begin{figure}[t]
   \centering
   \includegraphics[width=\columnwidth]{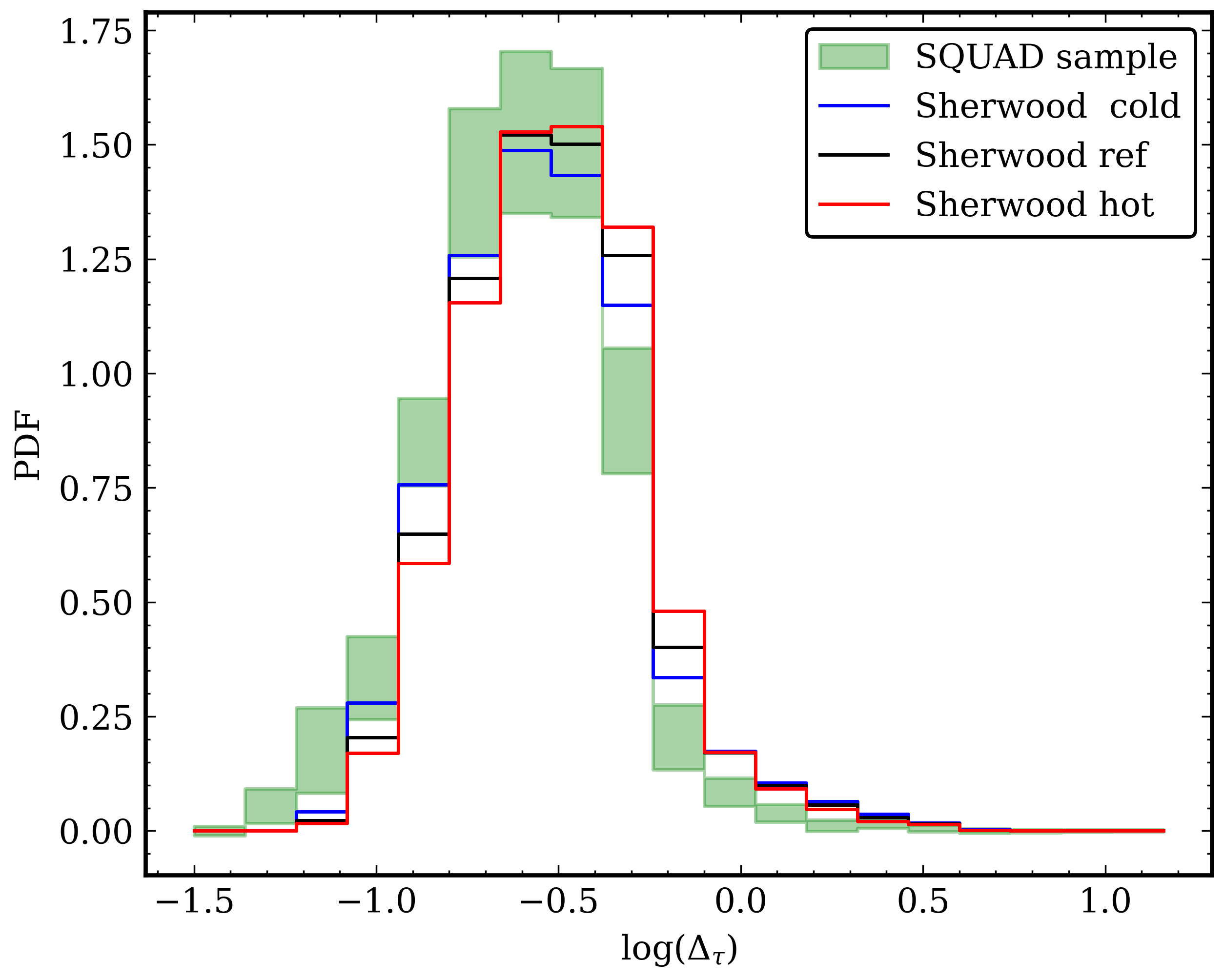}
      \caption{$\Delta_\tau$ PDF for the SQUAD sample using our Bayesian Neural network, the green shaded region represents the symmetric $1\sigma$ allowed region. The blue, black and red curves are, respectively, the PDFs computed over the whole simulation box for the CDM models of the \texttt{Sherwood-Relics} dataset.}
         \label{fig: squad pdf}
   \end{figure}

\subsection{Neural network performance validation on simulated data}
We first consider the validation split of the \texttt{Sherwood-Relics} suite, which has not been seen by the neural network during training. This split includes WDM mass runs and thermal runs, as well as the CDM reference run, with an equal number of skewers across all models. Since the validation split is $20\%$ of the total dataset, each run has 1000 randomly selected skewers. Globally, the neural network correctly recovers $85\%$ of the density pixels within $1\sigma$, and $97\%$ within $2\sigma$. The fact that more than $68\%$ of the pixels are recovered within the model's predicted $1\sigma$ uncertainty is an indication that the uncertainties are not gaussian; the recovered fractions behave in a more gaussian fraction when approaching the $2\sigma$ limit. As an illustration of a typical input-output to the trained network on \texttt{Sherwood-Relics} data, Figure \ref{fig: cdm_recovered} shows an example 20h$^{-1}$cMpc CDM skewer for the \texttt{Nyx} zre6 run. The top panel shows the input Lyman-$\alpha$ flux with added Gaussian noise (SNR $=50$) and an instrumental resolution of $6$ km\ s$^{-1}$. The bottom panel shows the machine learning prediction, with the true $\Delta_\tau$ field as a solid curve, and the mean prediction as a dashed curve. The blue region shows the $1\sigma$ uncertainty predicted by the Bayesian neural network. Note that the neural network has not been trained on any data generated using the \texttt{Nyx} code. We find that the mean prediction closely follows the ground truth in the regions where the absorption is not saturated and close to 0. That is, in the region where there is enough physical information, the neural network is able to accurately recover the underlying density field. By contrast, regions where the absorption is saturated (such as the region around 5 h$^{-1}$cMpc) are dominated by random noise. The network correctly identifies this and produces larger uncertainties than in the rest of the sightline.

In Figure \ref{fig: violin Nyx} we show the predictions residuals, as defined in equation \ref{eq:residuals}, for the 3 \texttt{Nyx} runs.  The corresponding percentages indicate the pixel fraction that is correctly recovered within $1\sigma$, which is $\geq 75\%$ for all 3 runs. Note that the performance is slightly degraded when predicting on \texttt{Nyx} as compared to the \texttt{Sherwood-Relics} validation data, which is expected since the former is not included in the training dataset. Accurate prediction on simulation Lyman-$\alpha$ using an alternative hydrodynamical code is a strong indication that the model is learning the physical relations that generate the data, rather than some simulation-specific artefact.

\begin{figure}[t]
   \centering
   \includegraphics[width=\columnwidth]{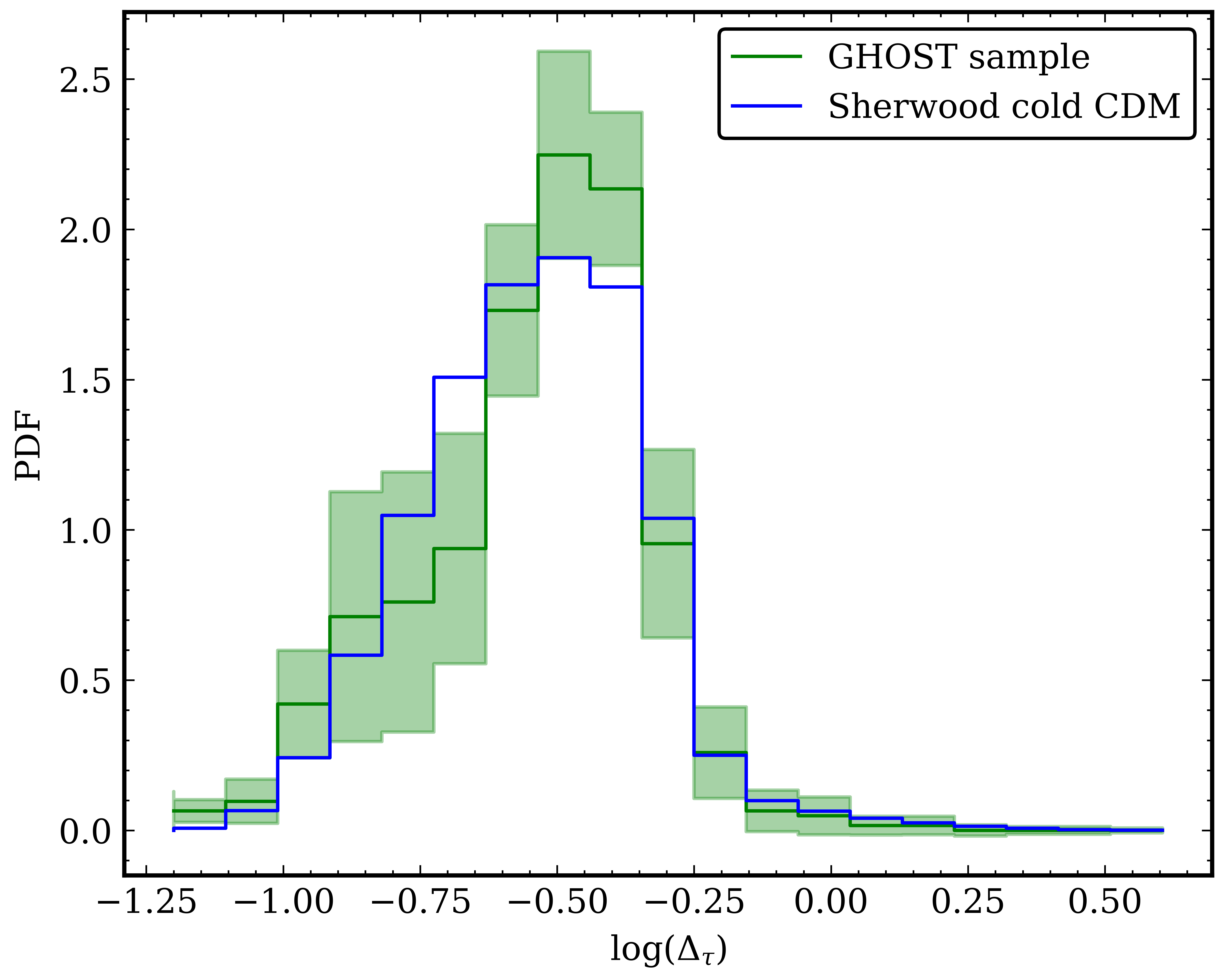}
      \caption{The $\Delta_\tau$ PDF (in green) reconstructed by using of Bayesian neural network for the two GHOST segments at $z=4.9$ from the quasar J0306+1853. We compare it to the best-fit \texttt{Sherwood-Relics} $\Delta_\tau$ PDF corresponding to the cold CDM run at $z=4.9$. We find a good agreement between the model and observed PDF, even when
       only two skewers are used at each redshift. }
         \label{fig: ghost fit pdf}
\end{figure}

\subsection{Constraints on multiple SQUAD DR1 spectra}
We now apply our machine learning inference pipeline described in Section \ref{sec: inference method} to the sample of 6 SQUAD DR1 targets listed in Table \ref{tab: squad dr1}. All sightlines are 20 h$^{-1}$cMpc and centred at $z=4.4$, but the noise levels depend on the target. To account for such variability when predicting, we retrain the model on the same \texttt{Sherwood-Relics} suite with the noise specifications corresponding to each observed sightline and then construct the $\Delta_\tau$ density field. Figure \ref{fig: squad_reconstruction} shows all 6 SQUAD sightlines in Table \ref{tab: squad dr1} together with the reconstructed $\Delta_\tau$ fields by our fiducial neural network trained at $z=4.4$ on the \texttt{Sherwood-Relics} suite. We compute the sightline $\Delta_\tau$ PDF from the total sample and fit it to the \texttt{Sherwood-Relics} PDFs obtained over all the skewers. Figure \ref{fig: squad pdf} shows the recovered PDF compared to the CDM PDF for all three \texttt{Sherwood-Relics} thermal models. 
We use equation \ref{eq:chi} to compute the $\chi^2$ metric between the observed data and the model PDFs for each WDM mass and thermal model. The $\chi^2$ metric is minimised by the \texttt{Sherwood-Relics} CDM cold model with $\chi^2_{\mathrm{min.}}\approx 10.5$. By inspecting the fit to the PDF, it appears that all \texttt{Sherwood-Relics} slightly under-predict the fraction of pixels with densities $\log(\Delta_\tau)<-0.5$, and that this tension is most alleviated with the cold models. We note that even colder models are unlikely to be physical, since $T_0$ in the cold model is already significantly below current measurements of the IGM temperature \citep{Gaikwad20, Gaikwad_2021}. Rather, this mismatch is more likely to be due to one of the other thermal parameters we are not currently taking into account, such as $\gamma$ or the patchiness of reionisation. 
The corresponding $2\sigma$ constraint on the WDM mass is then obtained using equation \ref{eq: confidence_region} and is $m_{\mathrm{WDM}} \gtrsim 3.8$ KeV at $2\sigma$ confidence.
In Figure \ref{fig: chi squad} we show the full $\chi^2$ metric as a function of the WDM particle mass for all three \texttt{Sherwood-Relics} thermal models (ref, cold and hot) for the SQUAD sample used in this section.

\subsection{Constraints from an individual GHOST spectrum}
We now centre our attention on the GHOST spectrum obtained from quasar J0306+1853 and introduced in Section \ref{sec: data}. For this sample, we have two Lyman-$\alpha$ skewers surrounding $z=4.9$ that we consider independent. To recover the $\Delta_\tau$ field the skewers, we retrain our Bayesian neural network with the specifications of each skewer, which have a different number of pixels. We use the snapshot from the \texttt{Sherwood-Relics} suite at $z=4.9$, a resolution of FWHM$=10$ km/s and SNR$=50$. Figure \ref{fig: ghost_reconstruction} shows both 20h$^{-1}$cMpc portions of the spectrum together with the recovered density fields. 

We apply our WDM inference pipeline to the two segments of the J0306+1853 spectrum. We combine the reconstructed $\Delta_\tau$ fields from the two segments redshifts and compute the $\chi^2$ metric in Equation \ref{eq:chi} between the reconstructed $\Delta_\tau$ fields and the $\texttt{Sherwood-Relics}$ model PDFs. 
Figure \ref{fig: chi ghost} shows the $\chi^2$ for all three \texttt{Sherwood-Relics} thermal models, as a function of the WDM particle mass. As expected, the $\chi^2$ is minimised for the CDM model within each thermal history, and the best-fit thermal model is again the \texttt{Sherwood-Relics} CDM cold run. 
The corresponding $2\sigma$ constraint, displayed in blue, on the WDM mass is $m_{\mathrm{WDM}} \gtrsim 2.2$ KeV at $2\sigma$ confidence. In Figure \ref{fig: ghost fit pdf} we show the $\Delta_\tau$ PDF (in green) reconstructed by using of Bayesian neural network for the two GHOST segments at $z=4.9$ from the quasar J0306+1853. We compare it to the best-fit \texttt{Sherwood-Relics} $\Delta_\tau$ PDF corresponding to the reference CDM run at $z=4.9$. We find a good agreement between the model and the observed PDF, even when
only two skewers are used. The constraints are slightly poorer than for the SQUAD sample, reflecting the issue with GHOST's focusing during the engineering run, which significantly impacted its effective resolution.

%


\section{Discussion}
Table \ref{tab: summary constraints} summarises and compares the current state-of-the-art $2\sigma$ lower bounds on $m_{\mathrm{WDM}}$ thermal relics constraints in the literature obtained from the Lyman-$\alpha$ power spectrum \citep{Villasenor_2023, Ir_i__2024}. The aforementioned efforts are based on a Bayesian inference framework using Markov Chain Monte Carlo techniques to compare the observed power spectrum with the one obtained from simulated data, typically limiting the analysis to $k$ scales $k\leq 0.1$. In contrast, in our work, we do the inference directly on the non-observable optical depth-weighted density field level. 
\cite{Ir_i__2024} demonstrate that the noise properties of the power-spectrum on small scales limit the obtainable constraints to roughly $m_\text{WDM} \gtrsim 4$, even if including the $k>0.1$ scales does formally tighten the signal. 
The resulting bounds produced in our work are comparably tight, but with the advantage of requiring substantially less observational data. For reference, in \cite{Ir_i__2024}, the constraints are obtained from 15 spectra measured across $4.0<z<5.2$, while our SQUAD DR1 data consists of 6  20h$^{-1}$cMpc skewers. This is $\sim 60$ times less observational data for a comparable WDM particle mass constraint to what is obtained from using the $k\leq 0.1$ scales of the power spectrum.


\begin{table}
      \caption[]{List of  current state-of-the-art $2\sigma$ lower bounds on $m_{\mathrm{WDM}}$ thermal relics constraints in the literature obtained from the Lyman-$\alpha$ power spectrum. Here, $k$ refers to the scales of the Lyman-$\alpha$ forest power spectrum, in units of s/km. We compare them to the results of this work, obtained doing inference directly at the density field level recovered by our Bayesian neural network. We also show the total path length
      used to obtain the constraints, computed by summing the comoving length of all the sightlines, as an indication of the amount of data required to obtain the results.}
         \label{tab: summary constraints}
     $$ 
         \begin{array}{p{0.4\linewidth}cc}
            \hline
            \noalign{\smallskip}
            Source &  m_{\mathrm{WDM}} \ [\mathrm{KeV}] & \mathrm{Path\ length} \ [\frac{\mathrm{cMpc}}{h}]  \\ 
            \noalign{\smallskip}
            \hline
            \noalign{\smallskip}
            \cite{Ir_i__2024}, $k \leq 0.1$& >4.1 & \sim 7300      \\
            \cite{Ir_i__2024}, all $k$& >5.7 & \sim 7300      \\
            \cite{Villasenor_2023} & >3.1 &\sim 7300     \\
            \noalign{\smallskip}
            \hline
            \noalign{\smallskip}
            This work (SQUAD DR1) & >3.8 &  120 \\
            This work (GHOST) & >2.2  & 40     \\

            \noalign{\smallskip}
            \hline
         \end{array}
     $$ 
   \end{table}

Since our method provides a full reconstruction of the density field, we can precisely quantify the WDM particle mass effect on the matter distribution. Traditional methods are limited to working with the Lyman-$\alpha$ flux field and use a single summary statistic at a time, since the optical depth-weighted density field is not directly accessible.

A limiting factor of this work is that we are only constraining a single parameter, $m_{\mathrm{WDM}}$, among a multitude that affects the statistics of the Lyman-$\alpha$ forest \citep{Weinberg_2003}. Further improvements built on top of our approach could potentially target to jointly constraint all parameters of interest: $m_{\mathrm{WDM}}$, the IGM thermal parameters, reionisation histories and cosmological parameters. 
The fact that both observational samples independently strongly prefer the \texttt{Sherwood-Relics} run with a cold IGM suggests that our method does have the potential to separate the effect of temperature and WDM smoothing, but our approach is not sufficient to jointly constrain those parameters. 
In contrast, the constraints obtained in \cite{Ir_i__2024} utilise a larger grid of simulations to marginalise over 7 parameters of interest ($m_{\mathrm{WDM}}$, $\tau_\mathrm{eff}$ $T_0$, $\gamma$, $u_0$, $\sigma_8$, and $n_s$). In addition, there is a concern that the mass resolution of the 20 Mpc/h simulation boxes of \texttt{Sherwood-Relics} is not sufficient to resolve the smallest scales where the effect of WDM is the strongest \cite{Ir_i__2024}.

\section{Conclusions}
In this work, we have built an inference pipeline at the density field level to constrain warm dark matter models that smooth the matter distribution in the Universe at a small scale. We have leveraged the \texttt{Sherwood-Relics} simulation suite, a set of high-resolution hydrodynamical simulations specially targeted at resolving the Lyman-$\alpha$ forest, to train a Bayesian neural network on a supervised regression task. Our machine learning model produces a posterior probability distribution for the target $\Delta_\tau$ optical depth-weighted density, is robust versus noise and accurately recovers $85\%$ of the fields in the \texttt{Sherwood-Relics} validation split. We have successfully tested the predictive capabilities of the trained neural network on a set of skewers obtained from the alternative simulation suite \texttt{Nyx}. Based on the recovered density fields, we construct a summary statistic, the Probability Distribution Function (PDF), and use it as the central object in the inference step. We fit the recovered density PDF from observed Lyman-$\alpha$ sightlines to the PDFs obtained from each thermal and WDM model in our simulation suite to constrain the $m_\mathrm{WDM}$ parameter. Applying this procedure to a set of 6 sightlines at redshift $z=4.4$ from the SQUAD DR1 sample, we find a lower bound of $m_{\mathrm{WDM}} \gtrsim 3.8$ KeV at $2\sigma$ confidence. We also test our inference pipeline on a slightly lower-resolution, high signal-to-noise ratio spectrum of J0306+1853, obtained using the GHOST instrument, finding that $m_{\mathrm{WDM}} \gtrsim 2.2$ KeV at $2\sigma$ confidence. Our results show that we can match state-of-the-art constraints using significantly less data, which highlights the potential of our approach. However, we also emphasize the necessity of including finer model grids in future works, enabling joint constraints of $m_\mathrm{WDM}$ and other relevant parameters such as thermal parameters of the IGM or cosmological parameters that can also affect the properties of the Lyman-$\alpha$ forest.

\begin{acknowledgements}
We warmly thank James Bolton and Vid Ir\v{s}i\v{c} for sharing the \texttt{Sherwood-Relics} simulations and for productive comments on the manuscript.

AA, SB, KP and BS are supported by the Deutsche Forschungsgemeinschaft (DFG) under Emmy Noether grant number BO 5771/1-1. EF is supported by the international Gemini Observatory, a program of NSF NOIRLab, which is managed by the Association of Universities for Research in Astronomy (AURA) under a cooperative agreement with the U.S.~National Science Foundation, on behalf of the Gemini partnership of Argentina, Brazil, Canada, Chile, the Republic of Korea, and the United States of America.

The Sherwood-Relics simulations were performed using the Joliot Curie supercomputer at Le Très Grand Centre de calcul (TGCC), the DiRAC Data Intensive service (CSD3) at the University of Cambridge, and the DiRAC Memory Intensive service Cosma6 at Durham University. CSD3 is managed by the University of Cambridge University Information Services on behalf of the STFC DiRAC HPC Facility (www.dirac.ac.uk). The DiRAC component of CSD3 at Cambridge was funded by BEIS, UKRI and STFC capital funding and STFC operations grants. Cosma6 is managed by the Institute for Computational Cosmology on behalf of the STFC DiRAC HPC Facility (www.dirac.ac.uk). The DiRAC service at Durham was funded by BEIS, UKRI and STFC capital funding, Durham University and STFC operations grants. DiRAC is part of the UKRI Digital Research Infrastructure. 

Based on observations obtained at the international Gemini Observatory, a program of NSF NOIRLab, which is managed by the Association of Universities for Research in Astronomy (AURA) under a cooperative agreement with the U.S. National Science Foundation on behalf of the Gemini Observatory partnership: the U.S. National Science Foundation (United States), National Research Council (Canada), Agencia Nacional de Investigaci\'{o}n y Desarrollo (Chile), Ministerio de Ciencia, Tecnolog\'{i}a e Innovaci\'{o}n (Argentina), Minist\'{e}rio da Ci\^{e}ncia, Tecnologia, Inova\c{c}\~{o}es e Comunica\c{c}\~{o}es (Brazil), and Korea Astronomy and Space Science Institute (Republic of Korea).

\end{acknowledgements}

%
%
\bibliographystyle{aa} 
\bibliography{biblio}

\begin{appendix}

\section{Robustness test on different resolutions}
In this section, we assess how training our neural network on a spectral resolution different from that of the observed data affects the $\Delta_\tau$ recovery results. We quantify this by considering our fiducial neural network trained at $z=4.4$ on the \texttt{Sherwood-Relics} suite with SNR $=50$ and a resolution of FWHM$=6$ km/s. We use this model with fixed weights and test its performance on the \texttt{Sherwood-Relics} validation split with a resolution of $4.8$ km/s and $7.2$ km/s, which corresponds to a resolution error of $20\%$. We also include for reference an extreme case with a resolution of $40$ km/s, to illustrate how this degrades to neural network's performance.

\begin{figure}[h!]
   \centering
   \includegraphics[width=\columnwidth]{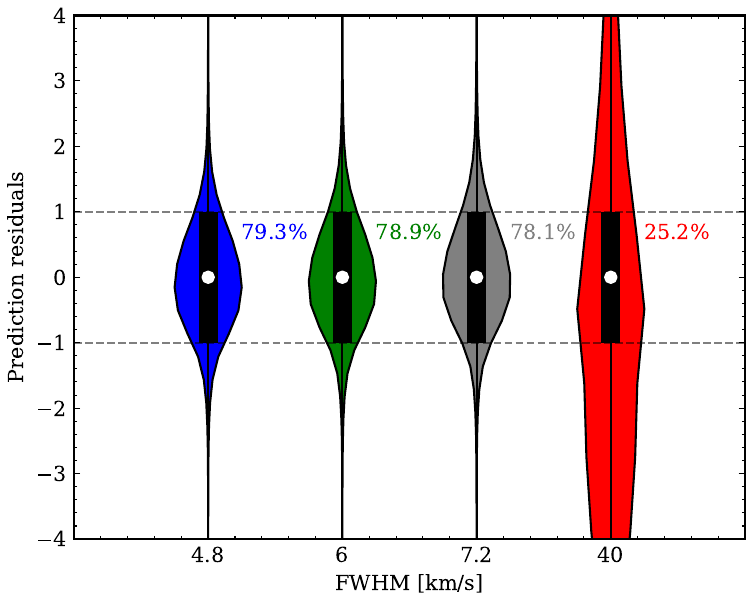}
      \caption{The distribution of the prediction residuals as defined in Equation \ref{eq:residuals} for all four different resolutions of $4.8, 6, 7.2$ and $40$ km/s when predicting on the \texttt{Sherwood-Relics} validation split with our fiducial neural network trained at $z=4.4$ with SNR $=50$ and FWHM$=6$ km/s.}
         \label{fig: test fwhm residuals}
\end{figure}

\begin{figure}[h!]
   \centering
   \includegraphics[width=\columnwidth]{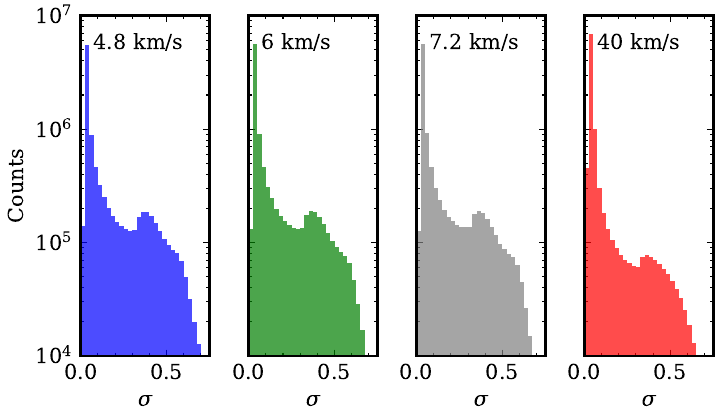}
      \caption{The distribution of the predicted uncertainties $\sigma$ for all four different resolutions of $4.8, 6, 7.2$ and $40$ km/s when predicting on the \texttt{Sherwood-Relics} validation split with our fiducial neural network trained at $z=4.4$ with SNR $=50$ and FWHM$=6$ km/s.}
         \label{fig: test fwhm sigmas}
\end{figure}

In Figure \ref{fig: test fwhm residuals} we show the distribution of the prediction residuals as defined in Equation \ref{eq:residuals} for all four different resolutions of $4.8, 6, 7.2$ and $40$ km/s. We find that changes of up to $20\%$ in the resolution of the observed data do not significantly change the performance of our network. In fact, a resolution increase in the observed data improves the performance by $\sim 0.4\%$, while a resolution decrease in the data degrades it by $\sim 0.8\%$. As expected, in the data with FWHM$=40$ km/s the small-scale features that are necessary to discriminate between WDM models are completely smoothed out and the performance is heavily affected. To assess if the neural network is palliating the loss of information by increasing the predicted uncertainties, we plot in Figure \ref{fig: test fwhm sigmas} the distribution of the uncertainties $\sigma$ predicted by the Bayesian neural network. Note how the distributions stay approximately constant when the model predicts on data with resolutions of $4.8, 6, 7.2$ km/s. There is no significant increase in the predicted uncertainties, and hence the neural network is robust against change in resolution of $\sim 20\%$.

\section{Reconstructed optical depth-weighted density fields}
\begin{figure*}[]
    \centering
    \includegraphics[width=0.7\textwidth]{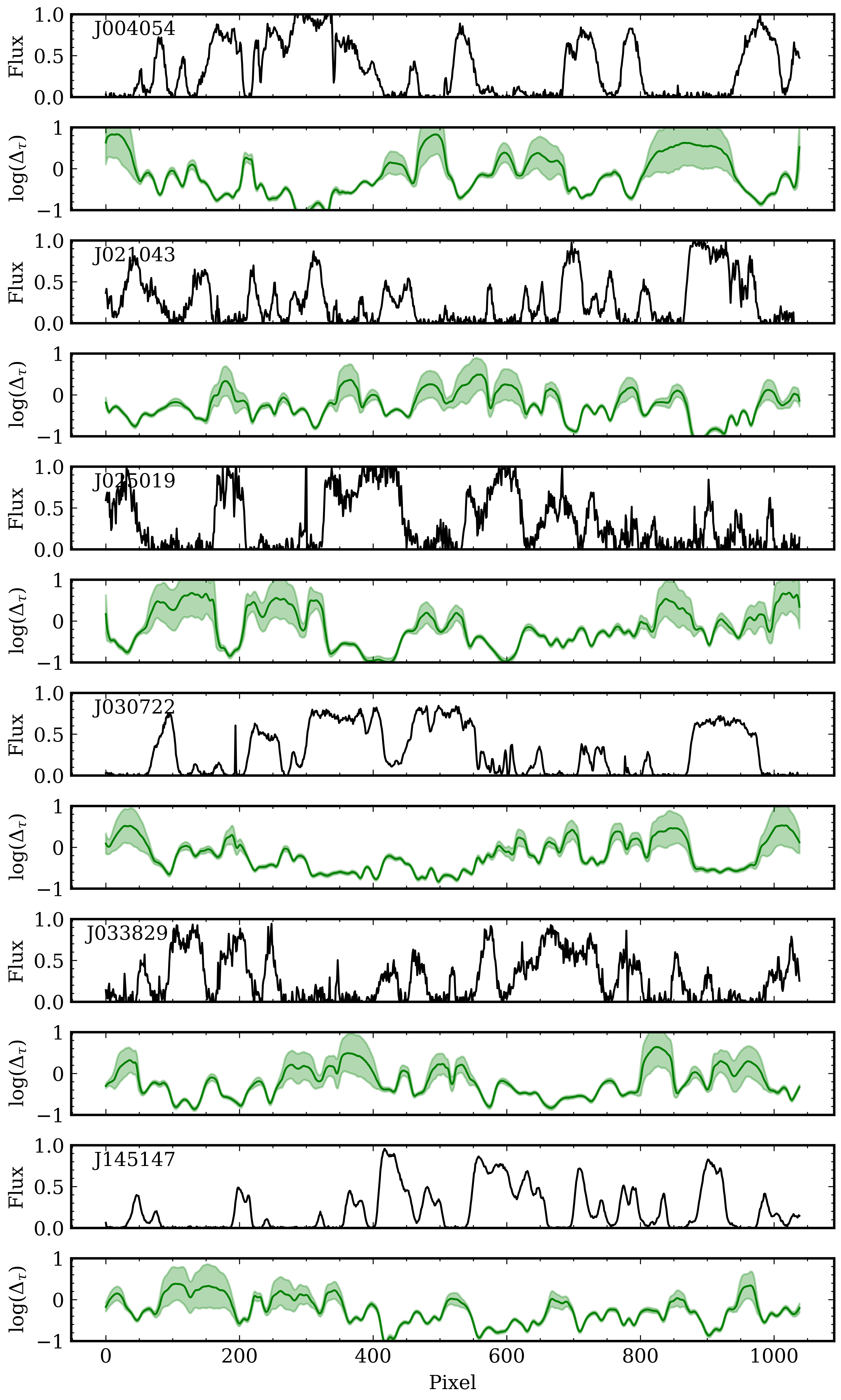}
    \caption{The SQUAD DR1 Lyman-$\alpha$ forest skewers from Table \ref{tab: squad dr1}, together with the reconstructed neutral hydrogen optical depth-weighted density $\Delta_\tau$ by our Bayesian neural network. All the skewers are 20h$^{-1}$cMpc in length (1039 pixels) and centred at redshift $z=4.4$.}
              \label{fig: squad_reconstruction}%
\end{figure*}

\begin{figure*}[]
    \centering
    \includegraphics[width=0.7\textwidth]{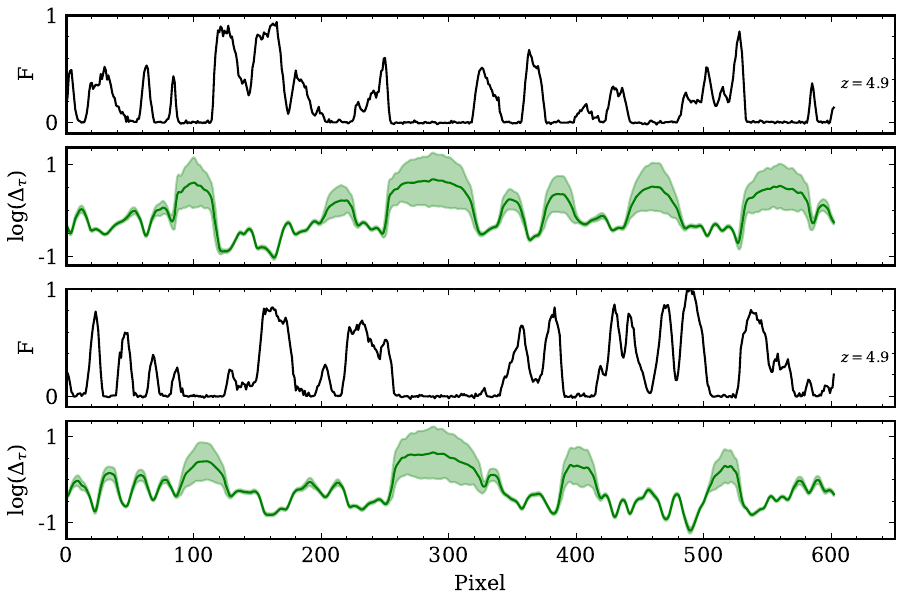}
    \caption{Both 20h$^{-1}$cMpc portions of the J0306+1853 spectrum together with the recovered density field by our Bayesian neural network. We split the original spectrum into 2 such skewers of length 20h$^{-1}$cMpc, and consider them independent. The top skewer terminates at $z=4.9$, while the second begins at that redshift. The number of pixels varies between the two segments.}
              \label{fig: ghost_reconstruction}%
\end{figure*}

We show in Figure \ref{fig: squad_reconstruction} and \ref{fig: ghost_reconstruction} the complete $\Delta_\tau$ reconstructed fields for the SQUAD DR1 and GHOST samples presented in Section \ref{sec: data}.

\section{Complements to the $\chi^2$ fits}
\begin{figure*}[]
    \centering
    \includegraphics[width=0.7\textwidth]{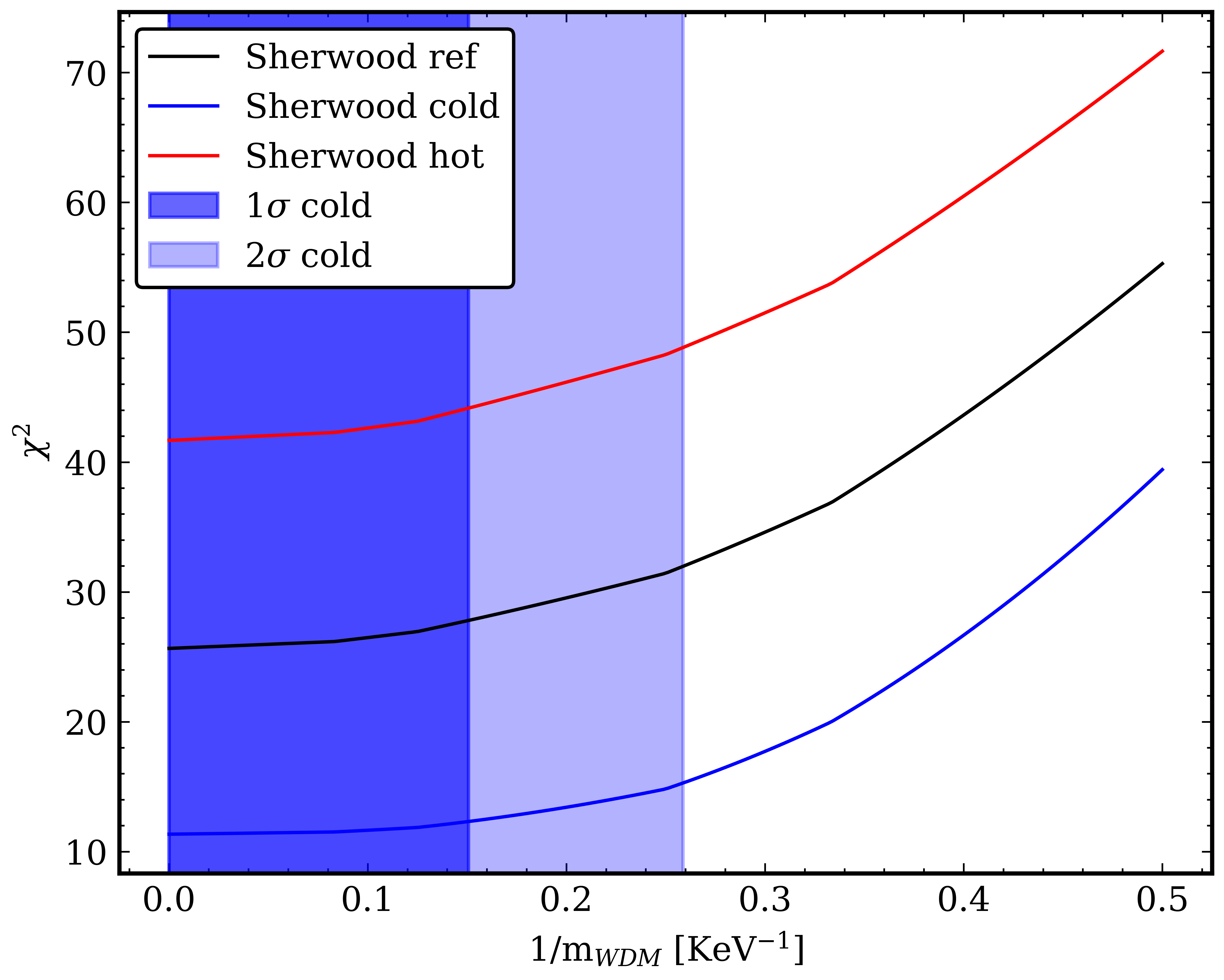}
    \caption{The $\chi^2$ metric for the SQUAD DR1 samples in Table \ref{tab: squad dr1} as a function of the WDM particle mass for all 3 thermal models in the \texttt{Sherwood-Relics} suite: ref, cold and hot. All skewers are centered at $z=4.4$. The length of the skewers is 20h$^{-1}$cMpc.}
              \label{fig: chi squad}%
\end{figure*}

\begin{figure*}[]
    \centering
    \includegraphics[width=0.7\textwidth]{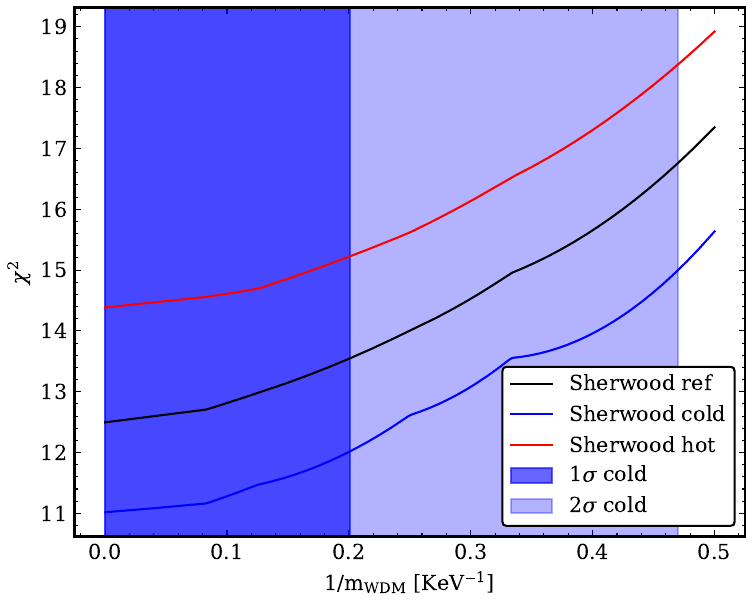}
    \caption{The $\chi^2$ metric for the GHOST samples from quasar J0306+1853 as a function of the WDM particle mass for all 3 thermal models in the \texttt{Sherwood-Relics} suite: ref, cold and hot. The sample spans is centred at $z=4.9$ and includes 2 skewers of length 20h$^{-1}$cMpc.}
              \label{fig: chi ghost}%
\end{figure*}

In Figure \ref{fig: chi squad} and Figure \ref{fig: chi ghost} we show the full $\chi^2$ fits as a function of the WDM particle mass for all 3 thermal models in the \texttt{Sherwood-Relics} suite: ref, cold and hot. See Table \ref{tab: Sherwood}.

\end{appendix}

\end{document}